\documentclass[12pt,draftclsnofoot,onecolumn]{IEEEtran} 
\usepackage{graphicx,cite,enumitem,}             		
			
\usepackage{amssymb,amsfonts,amsthm}   

\usepackage[cmex10]{amsmath}
\usepackage[unicode,colorlinks, linkcolor=black,anchorcolor=black,citecolor=black]{hyperref}

\def \tr {\text{tr}}

\usepackage{mathtools}

\usepackage{algorithm,caption}
\usepackage[noend]{algpseudocode}

\begin{document}

\title{Joint Receiver Design for Internet of Things}

\author{Kun~Wang \\ \today
}

\maketitle

\begin{abstract}
Internet of things (IoT) is an ever-growing network of objects that connect, collect and exchange data. 
To achieve the mission of connecting everything, physical layer communication is of indispensable importance. 
In this work, we propose a new receiver tailored for the characteristics of IoT communications. 
Specifically, our design is suitable for sporadic transmissions of small-to-medium sized packets in IoT applications.
With joint design in the new receiver, strong reliability is guaranteed and power saving is expected. 
\end{abstract}


\section{Introduction}

The Internet of Things (IoT) is a new technology introduced to 
facilitate people's daily lives and promote industry effectiveness by connecting a massive amount of smart devices \cite{Qualcomm2016Paving}.
As the foundation to a totally inter-connected world, IoT arouses great interest across academia and industry.
3GPP Release 13 has introduced a suite of key narrowband technologies optimizing for IoT, collectively referred to as LTE IoT.  
Specified by standard, LTE Cat-M1 (eMTC) \cite{Ericsson2014Further} aims at enhancing existing LTE networks -- 
coverage extension, UE complexity reduction, long battery lifetime and backward compatibility. 
Further, LTE Cat-NB1 (NB-IoT) scales down in cost and power for ultra-low-end IoT use cases \cite{wang2017primer}. 

A critical element for achieving these ultimate goals is the underlying physical communication link.
Rather than scale up in performance and mobility, the IoT communication networks are intended to be of 
low complexity and low power consumption.
Most IoT devices transmit small amount of data sporadically -- 
the maximum transport block size of downlink shared channel is 680 bits
and that of uplink shared channel is 1000 bits.
Our design of joint receiver aligns well with the requirements from IoT applications.
We target at FEC codewords of several hundred bits to a few thousand bits in length,
and through joint detection and decoding, the proposed receiver is able to outperform existing receivers in terms of block error rate,
thus saving HARQ rounds and ultimately consuming less power.

To fully take advantage of FEC code, we not only use FEC in decoding stage, but also make use of FEC code information in detection and demodulation.
Specifically, we will integrate the relaxed code constraints \cite{feldman2005using} to symbol detector in real/complex domain.
This set of code constraints have been widely used in our works \cite{wang2017galois}: 
space-time code as outer code is concatenated with LDPC code as inner code in \cite{wang2014joint,wang2015joint},
detection and demodulation with partial channel information is treated in \cite{wang2015diversity,wang2016diversity,wang2018unified},
multi-user scenario with different channel codes or different interleaving patterns are investigated in \cite{wang2016robust,wang2016fec}
and moreover, the asymmetry property of a class of LDPC code has been explored in \cite{wang2018semidefinite} to resolve the phase ambiguity.

\section{Decoding of binary FEC Codes on Real Field}
Among linear block codes, low-density parity-check (LDPC) code shows 
the capacity-approaching capability. 
An LDPC code $\mathcal{C}$ with parity check matrix $\mathbf{P}=[P_{i,j}]$ can
be represented by a Tanner graph $\mathcal{G} =
(\mathcal{V}, \mathcal{E})$.  Let $\mathcal{I} = \{1, 2, \ldots,
\emph{m}\}$ and ${\cal J} = \{1, 2, \ldots, {n}\}$,
respectively, be the row and column indices of $\mathbf{P}$. The
node set $\mathcal{V}$ can be partitioned into two disjoint node
subsets indexed by $\mathcal{I}$ and $\mathcal{J}$, known as the check nodes
and variable nodes, respectively. For each pair $(i , j) \in
\mathcal{I} \times \mathcal{J}$, there exists an edge (\emph{i},
\emph{j}) in $\mathcal{G}$ if and only if $P_{ij} = 1$. The index set of the
neighborhood of a check node $i \in \mathcal{I}$ is defined as
$\mathcal{N}_{i} := \{j \in \mathcal{J}: P_{i,j} = 1\}$. 
For each $i \in \mathcal{I}$, define the \emph{i}-th local code as
\vspace*{-1mm}
\begin{equation*}
\mathcal{C}_{i} = \{(c_{j})_{j \in \mathcal{J}}: \sum_{j
\in\mathcal{N}_{i}} P_{i,j}c_{j} = 0  \mbox{ in GF(2)}\}
\vspace*{-1mm}
\end{equation*}
where addition and multiplication are over \emph{GF}(2). 
Hence, a length-$n$ codeword
$\boldsymbol{c} \in \mathcal{C}$ if and only if 
$\boldsymbol{c} \in \mathcal{C}_i, \forall i \in \mathcal{I}$.
Therefore, decoding essentially needs to determine the most
likely binary vector $\mathbf{c}$ such that 
\vspace*{-1mm}
\begin{equation*}
\mathbf{P}\cdot \mathbf{c} = \mathbf{0}  \; \mbox{over GF(2)} 
\quad \mbox{or} \quad \Sigma_j P_{i,j}c_{j} = 0, \; \forall i\in {\cal I}.
\vspace*{-1mm}
\end{equation*}

Of interest are subsets $\mathcal{S} \subseteq \mathcal{N}_i$ that contain an even number of variable nodes; 
each such subset corresponds to a local codeword \cite{yang2006nonlinear}.
Let $\mathcal{E}_i \triangleq \{ \mathcal{S} \, | \, \mathcal{S} \subseteq \mathcal{N}_i \, \text{with} \, |\mathcal{S}| \, \text{even} \}$, and introduce auxiliary variable $v_{i,\mathcal{S}} \in \{0, 1\}$ to indicate the local codeword associated with $\mathcal{S}$. 
Since each parity check node can only be satisfied with one particular even-sized subset $\mathcal{S}$,  
the following equation must hold \cite{feldman2005using}
\vspace*{-1mm}
\begin{equation} \label{eq:code_v}
\sum_{\mathcal{S} \in \mathcal{E}_i} v_{i,\mathcal{S}} = 1, \; \forall i \in \mathcal{I}.
\vspace*{-1mm}
\end{equation}
Moreover, use $f_j \in \{0, 1\}$ to represent variable node $j$, indicating a bit value of 0 or 1.
The bit variables $f_j$'s must be consistent with each local codeword. Thus,
\vspace*{-1mm}
\begin{equation} \label{eq:code_f}
\sum_{\mathcal{S} \in \mathcal{E}_i: j \in \mathcal{S}} v_{i,\mathcal{S}} = f_j, \; \forall j \in \mathcal{N}_i, i \in \mathcal{I}.
\vspace*{-1mm}
\end{equation}

To see how these code constraints characterize a valid codeword at the $i$-th parity check, 
note that, according to constraint (\ref{eq:code_v}) and the fact that $v_{i,\mathcal{S}}$ takes integer values, 
we have $v_{i,\mathcal{S}'} = 1$ for some $\mathcal{S}'$ and $v_{i,\mathcal{S}''} = 0$ for all other $\mathcal{S}'' \neq \mathcal{S}'$,
where $\mathcal{S}', \mathcal{S}'' \in \mathcal{E}_i$.
Furthermore, from constraint (\ref{eq:code_f}), we have $f_j = 1$ for all $j \in \mathcal{S}'$ and 
$f_j = 0$ for all $j \in \mathcal{N}_i \backslash \mathcal{S}'$. 
Since $|\mathcal{S}'|$ is even-sized, the $i$-th parity check is satisfied.
Constraints (\ref{eq:code_v}) and (\ref{eq:code_f}) are enforced for every parity check. 
Together they define a valid codeword \cite{feldman2005using}. 
Notice that the constraint $v_{i,\mathcal{S}} \in \{0, 1\}$ would lead to integer programming, which is computationally expensive. 
Therefore, it is relaxed to $0 \leq v_{i,\mathcal{S}} \leq 1$. 
Meanwhile, constraint (\ref{eq:code_f}) guarantees that $0 \leq f_j \leq 1$.

The decoding constraints (\ref{eq:code_v}) and (\ref{eq:code_f}) use exponentially many variables $\{ v_{i,\mathcal{S}} \}$.
On the other hand, the constraints can be exponentially many, while with only $n$ variables $\{ f_i \}$.
This time, let $\mathcal{S} \triangleq \{ \mathcal{F} \, | \, \mathcal{F} \subseteq \mathcal{N}_i \, \text{with} \, |\mathcal{F}| \, \text{odd} \}$.
The fundamental polytope characterizing code property is captured by 
the following forbidden set (FS) constraints \cite{feldman2005using}
\begin{equation} \label{eq:parity_ineq}
\sum_{ i \in \mathcal{F} } f_i - \sum_{ i \in \mathcal{N}_i \backslash \mathcal{F}} f_i \leq |\mathcal{F}| - 1, \; \forall i \in \mathcal{I},
\forall \mathcal{F} \in \mathcal{S}
\end{equation}
plus the box constraints for bit variables
\begin{equation} \label{eq:box_ineq}
 0  \leq f_i \leq 1, \quad \forall i \in \mathcal{I}.
\end{equation}
In fact, if the variables $f_i$'s are zeros and ones, these constraints will be equivalent 
to the original binary parity-check constraints.
To see this, if parity check node $i$ fails to hold, there must be a subset of variable nodes
$\mathcal{F} \subseteq \mathcal{N}_i$ of odd cardinality such that all nodes in $\mathcal{F}$
have the value 1 and all those in $\mathcal{N}_i \backslash \mathcal{F}$ have value 0.
Clearly, the corresponding parity inequality in (\ref{eq:parity_ineq}) would forbid this situation.

\section{High-performance SDR receiver}
Maximum likelihood (ML) detection is known to be optimal in the sense of minimizing error probabilities. 
However, ML detection is exponentially complex, 
no matter exhaustive search or other search algorithm (e.g., sphere decoding) is used.
Recognizing the complexity, researchers showed great interest in the design of sub-optimal receivers.
Linear receivers, such as matched filtering (MF), zero-forcing (ZF) and minimum mean squared error (MMSE),
are widely used because of their simplicity. 
Besides linear receivers, more sophisticated receivers, for example, 
successive/parallel interference cancellation, are also used in practice.
However, the aforementioned receivers are far from optimal. 
In the recent decade or so, semi-definite relaxation (SDR), solved in polynomial time,
emerged as a new technique that achieves near-ML performance \cite{luo2010semidefinite}.

Consider an $N_t$-input $N_r$-output spatial multiplexing MIMO system with memoryless channel.
The baseband equivalent model of this system at time $k$ can be expressed as
\begin{equation} \label{eq:mimo_complex}
\mathbf{y}_k^c = \mathbf{H}_k^c \mathbf{s}_k^c + \mathbf{n}_k^c, \quad k = 1, \ldots, K,
\end{equation}
where $\mathbf{y}_k^c \in \mathcal{C}^{N_r \times 1}$ is the received signal,
$\mathbf{H}_k^c \in \mathcal{C}^{N_r \times N_t}$ denotes the MIMO channel matrix,
$\mathbf{s}_k^c \in \mathcal{C}^{N_t \times 1}$ is the transmitted signal, and 
$\mathbf{n}_k^c  \in \mathcal{C}^{N_r \times 1}$ is an additive Gaussian noise vector, 
each element of which is independent and follows $\mathcal{CN}(0, \sigma_n^2)$. 
In fact, besides modeling the point-to-point MIMO system, Eq.~(\ref{eq:mimo_complex}) can be also
used to depict frequency-selective system, multi-user system, etc. 
The only difference lies in the structure of channel matrix  $\mathbf{H}_k^c$.
To facilitate problem formulation, the complex-valued model is transformed to
a real-valued model by letting
$ \mathbf{y}_k = [\text{Re}\{ \mathbf{y}_k^c  \}^T \;  \text{Im}\{ \mathbf{y}_k^c  \}^T ]^T $,
$ \mathbf{s}_k =  [\text{Re}\{ \mathbf{s}_k^c  \}^T \; \text{Im}\{ \mathbf{s}_k^c  \}^T ]^T $,
$ \mathbf{n}_k = [\text{Re}\{ \mathbf{n}_k^c  \}^T \; \text{Im}\{ \mathbf{n}_k^c  \}^T ]^T$, 
and
\begin{equation*}
\mathbf{H}_k = 
\begin{bmatrix}
\text{Re}\{ \mathbf{H}_k^c  \}  & - \text{Im}\{ \mathbf{H}_k^c  \}\\
\text{Im}\{ \mathbf{H}_k^c  \} & \text{Re}\{ \mathbf{H}_k^c  \}
\end{bmatrix}.
\end{equation*}
Consequently, the transmission equation is given by
\begin{equation} \label{eq:mimo_real}
\mathbf{y}_k = \mathbf{H}_k \mathbf{s}_k + \mathbf{n}_k, \quad k = 1, \ldots, K.
\end{equation}
In this study, we choose capacity-approaching LDPC code for FEC purpose. 
Further, we assume the transmitted symbols are from Gray-labeled QPSK constellation,
i.e., $s_{k,i}^c \in \{ \pm1 \pm j \}$ for $k = 1, \ldots, K$ and $i = 1, \ldots, N_t$.
The codeword (on symbol level) is placed first along the spatial dimension
and then along the temporal dimension.  

The ML problem can then be equivalently written as the following QCQP
\begin{equation} \label{eq:qcqp}
\begin{aligned}
& \underset{\{\mathbf{s}_k, t_k\}}{\text{min.}}
& &  \sum_{k=1}^K 
\begin{bmatrix}
\mathbf{s}_k^T & t_k
\end{bmatrix}
\begin{bmatrix}
\mathbf{H}_k^T \mathbf{H}_k & \mathbf{H}_k^T \mathbf{y}_k  \\
-\mathbf{y}_k^T \mathbf{H}_k & || \mathbf{y}_k ||^2 
\end{bmatrix} 
\begin{bmatrix}
\mathbf{s}_k \\
t_k
\end{bmatrix} \\
& \text{s.t.}
& & t_k^2 = 1, \; s_{k,i}^2 = 1, \; k = 1, \ldots, K, i = 1, \ldots, 2N_t.
\end{aligned}
\end{equation}

This QCQP is non-convex because of its equality quadratic constraints. 
To solve it approximately via SDR, define the rank-1 semi-definite matrix $\mathbf{X}_k$ 
and the cost matrix $\mathbf{C}_k$
\begin{equation} \label{eq:rank1_matrix}
\mathbf{X}_k = 
\begin{bmatrix}
\mathbf{s}_k \\
t_k
\end{bmatrix}
\begin{bmatrix}
\mathbf{s}_k^T & t_k
\end{bmatrix}
=
\begin{bmatrix}
\mathbf{s}_k \mathbf{s}_k^T & t_k \mathbf{s}_k \\
t_k \mathbf{s}_k^T & t_k^2
\end{bmatrix},
\quad
\mathbf{C}_k = 
\begin{bmatrix}
\mathbf{H}_k^T \mathbf{H}_k & \mathbf{H}_k^T \mathbf{y}_k  \\
-\mathbf{y}_k^T \mathbf{H}_k & || \mathbf{y}_k ||^2
\end{bmatrix}.
\end{equation}
Based on the equality $\mathbf{v}^T\mathbf{Q}\mathbf{v} = \tr(\mathbf{v}^T\mathbf{Q}\mathbf{v}) = \tr(\mathbf{Q}\mathbf{v}\mathbf{v}^T)$,
the QCQP in Eq.~(\ref{eq:qcqp}) can be relaxed to SDR by removing the rank-1 constraint on $\mathbf{X}_k$.
\begin{equation} \label{eq:disjoint_sdr}
\begin{aligned}
& \underset{\{\mathbf{X}_k\}}{\text{min.}}
& &  \sum_{k=1}^K \tr(\mathbf{C}_k \mathbf{X}_k) \\
& \text{s.t.}
& & \tr(\mathbf{A}_i \mathbf{X}_k) = 1, \; k = 1, \ldots, K, i = 1, \ldots, 2N_t + 1, \\
& 
& & \mathbf{X}_k \succeq 0, \; k = 1, \ldots, K,
\end{aligned}
\end{equation}
where $\mathbf{A}_i$ is a zero matrix except that the $i$-th position on the diagonal is 1,
so $\mathbf{A}_i$ is used for extracting the $i$-th element on the diagonal of $\mathbf{X}_k$.
It is noted that $\mathbf{A}_i  \equiv \mathbf{A}_{i,k}, \forall k$; thus, the index $k$ is omitted
for $\mathbf{A}_{i,k}$ in Eq.~(\ref{eq:disjoint_sdr}). 

To anchor the FS constraints (\ref{eq:parity_ineq}) and (\ref{eq:box_ineq}) into the SDR formulation in Eq.~(\ref{eq:disjoint_sdr}),
one needs to connect the bit variables $f_n$'s with the matrix variables $\mathbf{X}_k$'s.
As stated in \cite{luo2010semidefinite}, if $(\mathbf{s}_k^*, t_k^*)$ is an optimal solution to (\ref{eq:disjoint_sdr}),
then the final solution should be $t_k^* \mathbf{s}_k^*$, where $t_k^*$ controls the sign of the symbol. 
As illustrated in Eq.~(\ref{eq:rank1_matrix}), the first $2 N_t$ elements of last column/row are exactly $t_k \mathbf{s}_k$.
We also note that the first $N_t$ elements correspond to the real parts of the transmitted symbols
and the next $N_t$ elements correspond to the imaginary parts. 
Hence, for Gray-labeled QPSK modulation, we have mapping constraints
for time instance $k = 1, \ldots, K$ as follows
\begin{equation} \label{eq:qpsk_gray}
\begin{split}
\tr(\mathbf{B}_i \mathbf{X}_k) & = 2 f_{2N_t(k-1)+2i-1} - 1, \; i = 1, \ldots, N_t, \\
\tr(\mathbf{B}_i \mathbf{X}_k) & = 1 - 2 f_{2N_t(k-1)+2i}, \; i = N_t+1, \ldots, 2N_t,
\end{split}
\end{equation}
where $\mathbf{B}_i$ is designed to extract the $i$-th element on the last row/column of $\mathbf{X}_k$.
Considering the symmetry of $\mathbf{X}_k$, we can define symmetric $\mathbf{B}_i$ as
\begin{equation}
\mathbf{B}_i = 
\begin{bmatrix}
0 & \ldots & \ldots & \ldots & 0 \\
\vdots   & \ddots &  & & \vdots  \\
\vdots   & & 0 &  & 1/2 \\
\vdots  & & & \ddots & \vdots \\
0 & \ldots & 1/2 & \ldots & 0
\end{bmatrix}, \; 1 \leq i \leq 2 N_t.
\end{equation}
The non-zero entries of $\mathbf{B}_i$ are the $i$-th element on the last row and the $i$-th element on the last column. 
With all components being ready, the joint SDR detector is assembled as follows
\begin{equation} \label{eq:joint_sdr}
\begin{aligned}
& \underset{\{\mathbf{X}_k, f_n\}}{\text{min.}}
& &  \sum_{k=1}^K \tr(\mathbf{C}_k \mathbf{X}_k) \\
& \text{s.t.}
& & \tr(\mathbf{A}_i \mathbf{X}_k) = 1, \, \mathbf{X}_k \succeq 0, \; k = 1, \ldots, K, i = 1, \ldots, 2N_t+1, \\
&
& & \text{tr}(\mathbf{B}_i \mathbf{X}_k) = 2 f_{2N_t(k-1)+2i-1} - 1, \; k = 1, \ldots, K, i = 1, \ldots, N_t, \\
&
& & \text{tr}(\mathbf{B}_i \mathbf{X}_k) = 1 - 2 f_{2N_t(k-1)+2i}, \; k = 1, \ldots, K, i = N_t+1, \ldots, 2N_t, \\
&
& & \sum_{ i \in \mathcal{F} } f_i - \sum_{ i \in \mathcal{N}_i \backslash \mathcal{F}} f_i \leq |\mathcal{F}| - 1, \; \forall i \in \mathcal{I}, \forall \mathcal{F} \in \mathcal{S}; \; 0  \leq f_i \leq 1, \; \forall i \in \mathcal{I}.
\end{aligned}
\end{equation}

Inspired by the turbo concept, we present an iterative SDR processing built upon the proposed joint SDR
in Eq.~(\ref{eq:joint_sdr}).
Without any \textit{a priori} information, we have all the bit variables $f_n$'s 
within the range $[0,1]$ in Eq.~(\ref{eq:box_ineq}).
After soft decoding, the \textit{a posteriori} LLRs contain more information 
(higher mutual information with the transmitted bits). 
It is known that the sign of an LLR determines the polarization of a bit, 
and the magnitude of an LLR implies its reliability. 
Therefore, we can select a certain bits of high reliability, 
and enforce stricter box constraints on those $f_i$'s in the next iteration.

\begin{algorithm}
\caption*{\textbf{Algorithm} Iterative SDR Processing} \label{alg:iter_sdr}
\begin{algorithmic}[1]
\State Obtain initial LLRs by solving the joint SDR of Eq.~(\ref{eq:joint_sdr}) and pass LLRs to SPA decoder.
\Repeat
\State Select a certain percentage of bits with high reliability according to the LLR magnitudes.
\State Impose stricter box constraints on the selected bits:
either $0 \leq f_n \leq \Pr[b_{n}=1 | \mathbf{Y}]$ or $\Pr[b_{n}=1 | \mathbf{Y}] \leq f_n \leq 1$ 
depending on the signs of LLRs.
\State Re-solve joint SDR with tightened box constraints and re-run SPA decoder.
\Until{Reach maximum number of iterations, or all parity checks are satisfied (examined by SPA decoder), 
or SDR is not feasible}.
\end{algorithmic}
\end{algorithm}

\begin{figure}[!tb]
\centering
\centerline{\includegraphics[width=9cm]{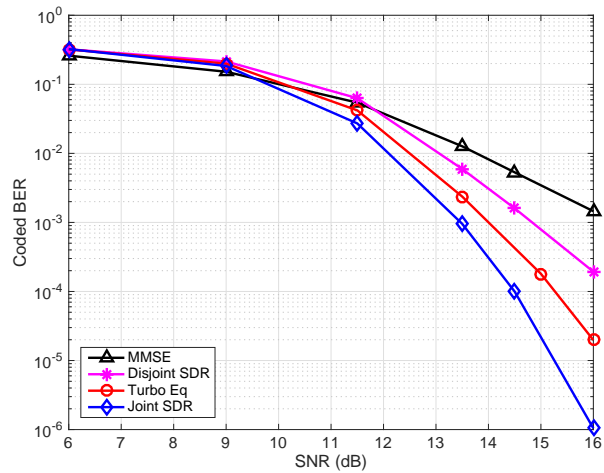}}
\caption{\small{BER comparisons of MMSE, disjoint SDR, turbo and joint SDR receivers.}}
\label{fig:all_comp}
\end{figure}

In Fig.~(\ref{fig:all_comp}), several receivers are compared in terms of coded BER.
We call the SDR formulation in Eq.~(\ref{eq:disjoint_sdr}) the disjoint SDR.
The BERs of turbo receiver and joint SDR receiver come from their respective last iteration.
It is clear that the proposed joint SDR achieves substantial gain over these existing receivers:
3 dB gain over MMSE at BER $10^{-3}$, 2 dB gain over disjoint SDR at BER $10^{-4}$ 
and 1 dB gain over turbo receiver at BER $10^{-5}$.

\section{Summary and Future Works}
This particular SDR formulation cannot be applied to arbitrary modulations;
each modulation should have its own SDR \cite{wang2018non,wang2018iterative,wang2018integrated}. 
For typical 16-QAM, several formulations were proposed in \cite{wiesel2005semidefinite, sidiropoulos2006semidefinite, yang2007mimo}.
However, they do not perform well in conjugation with FEC code constraints. 
SDR is an approximation to the non-convex QCQP, while there is another technique 
existing in the literature called reformulation linearization technique (RLT). 
As reported in \cite{anstreicher2009semidefinite}, the use of SDR and RLT constraints
together can produce bounds that are substantially better than either technique used alone.
So we should try RLT for MIMO detection of 16-QAM signaling with code constraints.
Besides higher-order modulation, we can also try to apply the technique to joint design with precoder \cite{wu2014cooperative,wu2015cooperative}.
In addition, the joint receiver design is useful to combat with RF imperfections \cite{wang2017phase}.

\ifCLASSOPTIONcaptionsoff
  \newpage
\fi



%

\bibliographystyle{IEEEtran}
\bibliography{IEEEabrv,mybibfile}

\end{document}